\documentclass{PoS}
\usepackage{graphicx}
\usepackage{amssymb}
\usepackage{amsmath}
\usepackage{multirow}
\usepackage{dcolumn}
\usepackage{nicefrac}

\newcommand{\sh}[1]{#1\hskip-7pt \diagup}

\title{Baryon properties from the covariant Faddeev equation}

\ShortTitle{Baryon properties from the covariant Faddeev equation}

\author{\speaker{Helios Sanchis-Alepuz}\\
        Institut f\"ur Physik, Karl-Franzens--Universit\"at Graz, Universit\"atsplatz 5, 8010 Graz, Austria\\
        E-mail: \email{helios.sanchis-alepuz@uni-graz.at}}

\author{Reinhard Alkofer\\
        Institut f\"ur Physik, Karl-Franzens--Universit\"at Graz, Universit\"atsplatz 5, 8010 Graz, Austria\\
        E-mail: \email{reinhard.alkofer@uni-graz.at}}

\author{Richard Williams\\
        Institut f\"ur Physik, Karl-Franzens--Universit\"at Graz, Universit\"atsplatz 5, 8010 Graz, Austria\\
        E-mail: \email{richard.williams@uni-graz.at}}

\abstract{A calculation of the masses and electromagnetic properties of the Delta and Omega baryon together with their evolution with the current quark mass is presented. Hereby a generalized Bethe-Salpeter approach with the interaction truncated to a dressed one-gluon exchange is employed. The model dependence is explored by investigating two forms for the dressed gluon exchange.}

\FullConference{Sixth International Conference on Quarks and Nuclear Physics\\
		 April 16-20, 2012\\
		 Ecole Polytechnique, Palaiseau,  Paris}

\begin{document}

\section{Introduction}
A covariant description of relativistic two- and three-body bound states is provided by the (generalised) Bethe-Salpeter equations. Until recently quark-diquark calculations of baryons were standard~\cite{hep-ph/9705267,nucl-th/9805054,nucl-th/9907120,arXiv:0812.1665,arXiv:1008.3184}, whereas today calculations have reached parity with meson studies in the form of a more intricate three-body description~\cite{arXiv:0912.2246,arXiv:1104.4505,arXiv:1109.0199,heliosthesis} at the level of the Rainbow-Ladder (RL) approximation. In the meantime, meson studies have made progress beyond RL~\cite{arXiv:0808.3372,arXiv:0905.2291,arXiv:0903.5461,Williams:2009wx}.

In the case of baryons, only one RL interaction has typically been tested, known as the Maris-Tandy model~\cite{nucl-th/9708029,nucl-th/9905056}.
This dominance is well-earned since it performs well phenomenologically. However, it does not draw on the plethora of information we now know about QCD Green's functions and so we choose here to investigate in addition another effective interaction.

With the truncation and effective interactions thus chosen, we shall proceed to calculate the appropriate meson and baryon masses, together with a calculation of the electromagnetic properties of the Delta baryon in Rainbow-Ladder approximation. Here we provide a summary of the results; for details we refer to \cite{heliosthesis} and references therein.

\section{Framework}
We start with the DSE for the quark propagator,
\begin{equation}
	S^{-1}(p) = Z_2 \,S_0^{-1} + g^2 \,Z_{1f}\int \frac{d^4k}{\left( 2\pi
	\right)^4} \gamma^\mu S(k) \,\Gamma^\nu(k,p) \,D_{\mu\nu}(k-p)\,\, .
	\label{eqn:quark_DSE}
\end{equation}
Here $S^{-1}(p) = A(p^2) \left(  i\sh{p} + M(p^2) \right)$ is the inverse quark propagator, with $S_0^{-1}(p)$ its bare counterpart. The quark wave-function renormalisation is $1/A(p^2)$   and $M(p^2)$
the quark mass function. $Z_2$ and $Z_{1f}$ are renormalisation constants of the quark propagator and quark-gluon vertex respectively. In Landau gauge, $D_{\mu\nu}$ is just the transverse projector $T_{\mu\nu}(q) = \delta_{\mu\nu} - q_\mu q_\nu/q^2$ multiplied by the
 gluon dressing function $Z(q^2)/q^2$. We combine $D_{\mu\nu}$ with  $\Gamma^\nu(k,p)$ such that
\begin{equation}
Z_{1f}\,\frac{g^2}{4\pi} \,D_{\mu\nu}(q) \,\Gamma_\nu(k,p)
  = Z_2^2 \, T_{\mu\nu}(q) \,\frac{\alpha_{\rm eff}(q^2)}{q^2}\,\gamma_\nu\,.
\end{equation}
Here $\alpha_{\rm eff}(q^2)$ is an effective interaction subsuming non-perturbative features of the gluon propagator and the quark-gluon vertex, and $Z_2^2$ follows from Slavnov-Taylor identities.  Consideration of chiral symmetry leads to the two-body kernel in RL approximation,
\begin{equation}
	K^{\textrm{2-body}}= 4\pi \,Z_2^2 \,\frac{\alpha_{\rm eff}(q^2)}{q^2}\,
	T_{\mu\nu}(q)\,\gamma^\mu \otimes \gamma^\nu\,\,.
	\label{eqn:ladder}
\end{equation}
For the baryon, the three-body kernel $K^{\textrm{3-body}}$ is decomposed into an irreducible three-quark
contribution and the sum of permuted two-body
kernels $K^{\textrm{2-body}}$.

In this paper we compare two interactions. The first is the Maris-Tandy (MT) model mentioned above. The second, which we will refer to as (AFW),  has been proposed in reference~\cite{arXiv:0804.3478}, combining the gluon dressing~\cite{hep-ph/0309077} with a model for the quark-gluon vertex~\cite{arXiv:0804.3042}. For more details, see \cite{SanchisAlepuz:2011aa}.

\section{Hadron Masses}

\begin{figure}[t]
\centering
\includegraphics[height=0.34\textheight,clip]{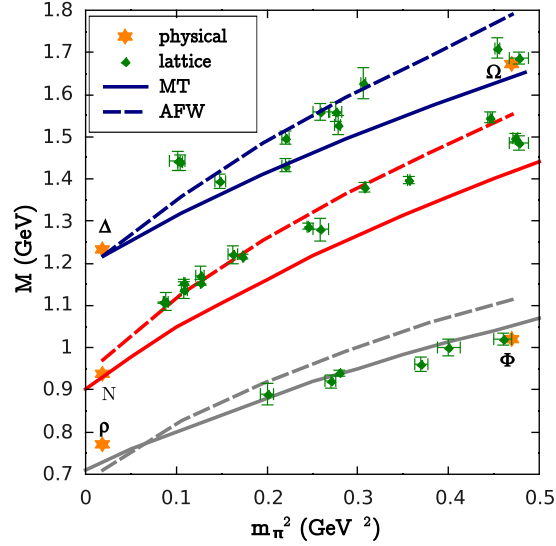}
\caption{Evolution of $\rho$, $N$ and $\Delta$ masses
with $m_\pi^2$ for MT and AFW models. We compare
to lattice data; see~\cite{arXiv:1104.4505,arXiv:1109.0199} for references.}
\label{fig:masses}
\end{figure}

Both models reproduce the pattern of dynamical chiral symmetry breaking (DCSB) and related observables. This is also seen for light states such as the rho, nucleon and delta where both models describe the data well, as seen in Fig.~\ref{fig:masses}. This is not surprising as they both exhibit similar features at the important momentum region of $\sim 1$~GeV, relevant for DCSB. As the quark masses evolve to heavier values, a greater deviation is seen between the two models although qualitative agreement with lattice and/or experiment is always within $\sim 10$\%. This implies that we have a qualitative model independence within RL. We collect results in Table \ref{tab:results}.

\begin{table*}[h]
\begin{center}
\begin{tabular}{c@{\!\;\;}l|ccc} 
\multicolumn{2}{c|}{$J^{PC}=0^{-+}$}  & MT         & AFW         & exp. \\ \hline
$n\overline{n}$ &$(\pi)$               & 0.140\dag  & 0.139\dag   & 0.138 \\
$n\overline{s}$ &$(K)$                 & 0.496\dag  & 0.497\dag   & 0.496 \\
$c\overline{c}$ &$(\eta_c)$            & 2.979\dag  & 2.980\dag   & 2.980 \\
$b\overline{b}$ &$(\eta_b)$            & 9.388\dag  & 9.390\dag   & 9.391 \\ \hline
\multicolumn{2}{c|}{$J^{PC}=1^{--}$}  & MT         & AFW         & exp. \\ \hline
$n\overline{n}$ &$(\rho)$              & 0.743      & 0.710       & 0.775 \\
$n\overline{s}$ &$(K^\star)$           & 0.942      & 0.961       & 0.892 \\
$s\overline{s}$ &$(\phi)$              & 1.075      & 1.114       & 1.020 \\
$c\overline{c}$ &$(J/\psi)$            & 3.163      & 3.302       & 3.097 \\
$b\overline{b}$ &$(\Upsilon)$          & 9.466      & 9.621       & 9.460 \\ \hline
\end{tabular} \hspace{5mm}
\begin{tabular}{c|cccc} 
                    & MT    & AFW   & exp. &   \\ \hline
$N$                 & 0.94  & 0.97  & 0.94 &      \\
$\Delta$            & 1.26  & 1.22  & 1.23 &      \\
$\Omega$            & 1.72  & 1.80  & 1.67 &       \\ \hline
                    & MT    & AFW   & lattice  & pNRQCD  \\ \hline
$\Omega_{ccc}$      & 4.4   & 4.9   & 4.7      & 4.9(0.25)        \\
$\Omega_{bbb}$      & 13.7  & 13.8  & 14.4     & 14.5(0.25)        \\
\end{tabular}
\caption{Meson and baryon masses (GeV) for both interactions (fitted values marked $\dag$), compared to experiment. Heavy-Omega baryons are not yet observed thus we compare
to lattice ~\cite{hep-lat/0501021,arXiv:1010.0889} and pNRQCD~\cite{arXiv:1111.7087}.}\label{tab:results}
\end{center}
\end{table*}

For the heavy quarks, see also Table \ref{tab:results}, we see that the charm quarks exhibit similar trends with AFW giving heavier masses than MT. Unexpectedly for bottom quarks the calculated masses coincide. We see that the $\Upsilon$ mass is well reproduced however the $\Omega_{bbb}$ is less so. One may speculate here as to the relevance of three-body interactions. To make precise statements one should fit the models to the heavy quark sector where corrections beyond RL are suppressed, and then study the evolution to light quarks. We remark that the majority of RL models already capture beyond-RL effects at u/d quark masses in the parameterisation of the interaction.

\section{Electromagnetic Form Factors of the $\Delta$}
The current for a spin-\nicefrac{3}{2} particle is characterised by four form factors $F_1$\ldots$F_4$,
\begin{eqnarray}
J^{\mu,\alpha\beta}(P,Q) &=& \mathbb{P}^{\alpha\alpha\prime}(P_f)
\bigg[ \left( \left( F_1 + F_2\right) i\gamma^\mu -F_2 \frac{P^\mu}{M}\right) \delta^{\alpha\prime\beta\prime}
\nonumber \\ &&\hspace{1.3cm}+\left( \left( F_3 + F_4\right) i \gamma^\mu -F_4 \frac{P^\mu}{M} \right) 
\frac{Q^{\alpha\prime}Q^{\beta\prime}}{4M^2} \bigg] \mathbb{P}^{\beta\prime\beta}(P_i)\;\;,
\end{eqnarray}
where $\mathbb{P}$ is the Rarita-Schwinger projector, $P_i$ and $P_f$ are the initial and final baryon total momenta, respectively, and $Q=P_f-P_i$. These form factors can be related to the usual electric monopole $G_{E_0}$, magnetic dipole $G_{M_1}$, electric quadrupole $G_{E_2}$ and magnetic octupole $G_{M_3}$. We use the \emph{gauging of equations} technique, \cite{Kvinikhidze:1998xn,Kvinikhidze:1999xp} to couple our baryon to an external EM field such that gauge symmetry is preserved.  Definitions and conventions for this and further details can be found in \cite{Nicmorus:2010sd,heliosthesis}.

\begin{figure}[h]
\centering
\includegraphics[height=0.48\textheight,clip]{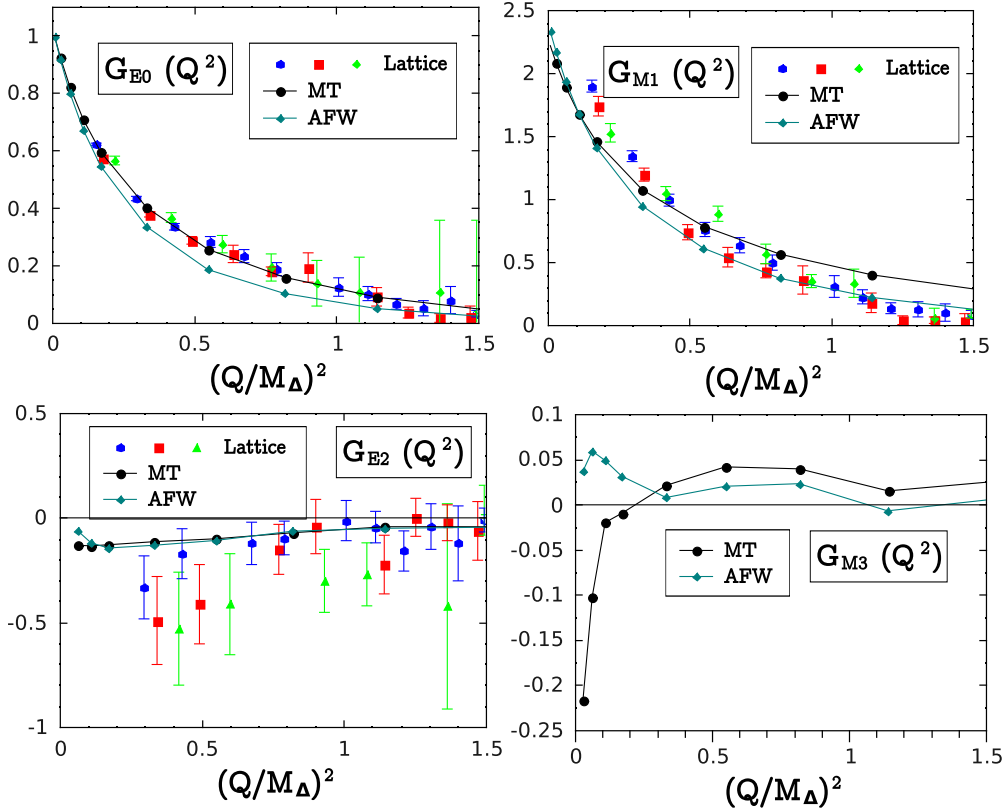}
\caption{(color online) EM form factors for the $\Delta^+$ for the MT and AFW models. The results are compared to lattice data \cite{Alexandrou:2009hs} for dynamical Wilson fermions at $m_\pi=384$ MeV (green), $m_\pi=509$ MeV (red) and $m_\pi=691$ MeV (blue).}
\label{fig:form}
\end{figure}

In Fig.~\ref{fig:form} we show results for the $\Delta^+$ form factors. Note that owing to isospin symmetry, for the $\Delta^{++}$, $\Delta^0$ and $\Delta^-$ their form-factors differ only by a factor corresponding to their charge.  Due to the short lifetime of the Delta resonance it is  difficult to study it experimentally, and EM properties are restricted to the $\Delta^{++}$ and $\Delta^{+}$ magnetic dipoles with small statistics. Thus we compare with a lattice calculation using dynamical Wilson fermions at different pion masses
\cite{Alexandrou:2009hs}. These presently suffer from large errors, especially for the electric quadrupole form factor and give no information on the magnetic octupole. 

We see at large $Q$ good agreement with the lattice where the quark core is probed. At smaller values the pion cloud is not captured by our model and so deviations are expected. Furthermore in this region, the electric quadrupole and magnetic octupole feature numerical inaccuracies due to cancellations enhanced by $1/Q^4$ and $1/Q^6$ terms, as discussed in~\cite{heliosthesis}. Improvement of this is in progress.

\begin{table}[ht]
\centering
\begin{tabular}{c|cc|ccc|c}

	& F-MT & F-AFW & DW1 & DW2 & DW3 & Exp. \\
\hline
\hline
$\left< r_{E_0}^2\right>$(fm$^2$) & 0.67 & 0.60 & 0.373 (21) & 0.353 (12) & 0.279 (6) & \\
$G_{M_1}(0)$                      & 2.22 & 2.33 & 2.35  (16) & 2.68  (13) & 2.589 (78) & 3.54$^{+4.50}_{-4.72}$ \\
\end{tabular}
\caption{Comparison of results for the charge radius $\left<r_{E_0}^2\right>$ and for $G_{M_1}(0)$. We compare our two model calculations to the lattice at $m_\pi=384$ MeV (DW1), $m_\pi=509$ MeV (DW2) and $m_\pi=691$ MeV (DW3) \cite{Alexandrou:2009hs,Alexandrou:2009nj}. Where available we also show experiment~\cite{Nakamura:2010,Kotulla:2002cg}.\label{tab:radius}}
\end{table}

In Table~\ref{tab:radius} we show the computed results for the charge radius of the $\Delta^+$ baryon. For both models, our results appear considerably higher than those of the lattice. A possible explanation is the pion-mass
dependence of the charge radius, which grows as the pion mass approaches the physical value from above. Moreover $\chi$PT shows than when the $\Delta\rightarrow N\pi$ decay channel opens the charge radius changes abruptly to a lower value~\cite{Ledwig:2011cx}. Since in our calculation we have no mechanism for the Delta to decay, our larger result is not unreasonable.

\section{Summary and Outlook}

We have shown results for the electromagnetic form factors of the Delta baryon using a three-body covariant Bethe-Salpeter equation and two different models for the effective interaction. Our results show good agreement with lattice data, especially at large photon momentum, and feature a qualitative model independence. Calculations of the electromagnetic form factors for the Omega as well as refinements in the numerical techniques are in progress.

\acknowledgments
We thank G. Eichmann and C.~Fischer for useful discussions. 
This work was supported by the Austrian Science Fund FWF under Projects No. P20592-N16 and No. M1333-N16 and the Doctoral Program W1203 (Doctoral Program ``Hadrons in vacuum, nuclei and stars'').

\end{document}